# Breakdown Current Density in BN-Capped Quasi-1D TaSe$_3$ Metallic Nanowires: Prospects of Interconnect Applications

Maxim A. Stolyarov[1][*], Guanxiong Liu[1][*], Matthew A. Bloodgood[2], Ece Aytan[1], Chenglong Jiang[1], Rameez Samnakay[1], Tina T. Salguero[2], Denis L. Nika[1,3], Krassimir N. Bozhilov[4], and Alexander A. Balandin[1,†]

[1]Nano-Device Laboratory (NDL) and Phonon Optimized Engineered Materials (POEM) Center, Department of Electrical and Computer Engineering, University of California – Riverside, Riverside, California 92521 USA

[2]Department of Chemistry, University of Georgia, Athens, Georgia 30602 USA

[3]E. Pokatilov Laboratory of Physics and Engineering of Nanomaterials, Department of Physics and Engineering, Moldova State University, Chisinau, MD-2009 Republic of Moldova

[4]Central Facility for Advanced Microscopy and Microanalysis, University of California – Riverside, Riverside, California 92521 USA

---

[*] These authors contributed equally to the project.
[†] Corresponding author (A.A.B.): balandin@ece.ucr.edu

**Abstract**

We report results of investigation of the current-carrying capacity of nanowires made from the quasi-1D *van der Waals* metal tantalum triselenide capped with quasi-2D boron nitride. The chemical vapor transport method followed by chemical and mechanical exfoliation were used to fabricate mm-long TaSe$_3$ wires with lateral dimensions in the 20 to 70 nm range. Electrical measurements establish that TaSe$_3$/*h*-BN nanowire heterostructures have a breakdown current density exceeding 10 MA/cm$^2$ — an *order-of-magnitude* higher than that in copper. Some devices exhibited an intriguing step-like breakdown, which can be explained by the *atomic thread* bundle structure of the nanowires. The quasi-1D single crystal nature of TaSe$_3$ results in low surface roughness and the absence of grain boundaries; these features potentially can enable the downscaling of these wires to lateral dimensions in the few-nm range. These results suggest that quasi-1D van der Waals metals have potential for applications in the ultimately downscaled local interconnects.

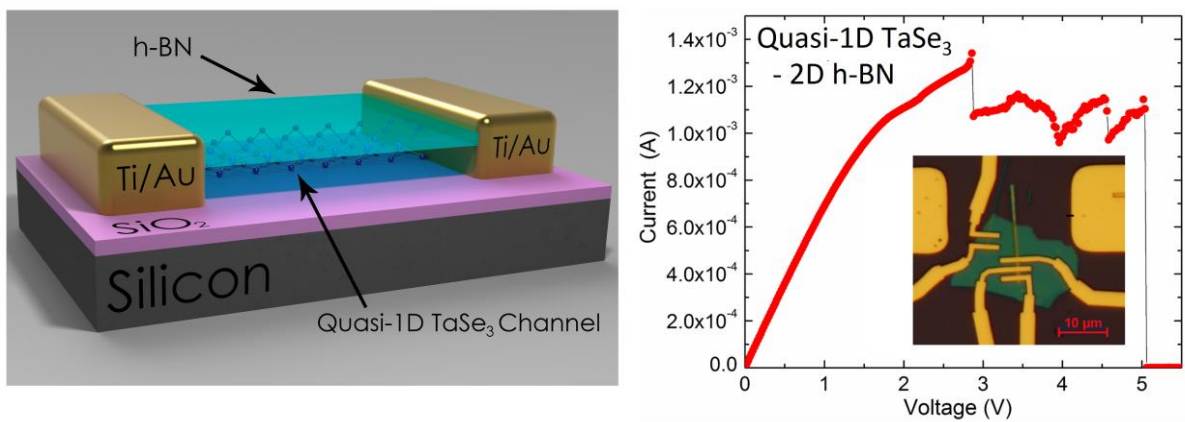

**Contents Image**

Recently discovered unique electrical [1, 2], thermal [3, 4] and optical [5-7] properties of graphene has stimulated the search for other two-dimensional (2D) atomic crystals and heterostructures with properties distinct from the corresponding bulk materials [8-16]. A large number of 2D materials belong to the family of transition metal chalcogenides that contains weak van der Waals bonding between structural units [17, 18]. Most research up to now has utilized dichalcogenide compositions $MX_2$ (where M = Mo, W, and other transition metals; X = S, Se, Te), such as $MoS_2$, $TaSe_2$, $TaS_2$ and others [19-27], distinguished by 2D layers. Another class of transition metal chalcogenides, the trichalcogenides, has quasi-1D crystalline structures. Examples include $TiS_3$, $TaSe_3$, $TaS_3$ and $NbSe_3$ [28-31]. The crystal structure views of monoclinic tantalum triselenide ($TaSe_3$) in Figure 1 (a) illustrate the main features of this material. Trigonal prismatic $TaSe_3$ forms continuous chains that extend along the b axis, leading to fiber- or needle-like crystals with anisotropic metallic properties. These chains are arranged in corrugated bilayer sheets, indicated by blue and yellow rows of prisms in Figure 1 (a). The van der Waals bonding between sheets is weak, which allows $MX_3$ materials to be exfoliated by the same mechanical methods as $MX_2$ materials.

A study of the current-carrying capacity of exfoliated $MX_3$ materials is interesting from both fundamental science and practical applications points of view. A possibility of achieving very high breakdown current densities in quasi-1D metallic conductors may have implications for electronic industry. Indeed, continuous downscaling of the silicon (Si) complementary metal-oxide-semiconductor (CMOS) technology results in increasing current densities in the copper (Cu) interconnects. According to the International Technology Roadmap for Semiconductors (ITRS), the present level of the current density, ~1.8 MA/cm$^2$ at the half-pitch width of 28.5 nm will increase to ~5.35 MA/cm$^2$ at the width of 7 nm [32]. There is no existing technology with the breakdown current density, $J_B$, high enough to sustain such currents. The ITRS projections indicate that the layer thicknesses will decrease from 57.0 nm currently in 2016 to 15.4 nm by 2028, while the interconnect cross-sectional area will scale down from 1624.5 nm$^2$ to 107.8 nm$^2$ over the same period of time. Scaling deep to the nanoscale range presents problems for conventional metals due to their polycrystalline structure, surface roughness and increased electrical resistivity owing to the electron–boundary scattering [33]. These factors motivate the search for alternative materials, which can complement Cu in selected areas, e.g. ultimately downscaled local interconnects.

In this Letter, we report results describing the current-carrying capacity of nanowires made from metallic TaSe$_3$ capped with hexagonal boron nitride (*h*-BN). This work establishes that quasi-1D TaSe$_3$ nanowires have the breakdown current density exceeding $J_B$~10 MA/cm$^2$, which is an order-of-magnitude larger than that in Cu nanowires with diffusion barriers. In view of the promising current-carrying capacity of such quasi-1D/quasi-2D heterostructures and the possibility of ultimately downscaling their cross-sectional areas, we discuss the prospects of using quasi-1D metals as interconnects.

High-quality TaSe$_3$ crystals were prepared by the chemical vapor transport (CVT) method using iodine (see Methods). The morphology, crystalline structure, and phase purity of these crystals were characterized by scanning electron microscopy (SEM), electron dispersive spectroscopy (EDS), powder X-ray diffraction (XRD), and electron probe micro analysis (EPMA). Figure 1 (b) shows an SEM image of a representative fibrous TaSe$_3$ crystal ~225 µm wide and >10 mm long. The powder XRD pattern in Figure 1 (d) matches literature data for standard monoclinic TaSe$_3$ (P2$_1$/m) [31] , albeit several peaks (marked with *) exhibit enhanced intensity due to the preferred orientation of particles within the powdered sample. Furthermore, both EDS and EPMA analyses (Supplemental Information) show that the stoichiometry of CVT-grown crystals is consistent with TaSe$_3$.

[Figure 1 (a-e): TaSe$_3$ material characterization.]

The crystals were subjected to successive chemical and mechanical exfoliation. The chemical exfoliation involved low-power ultrasonic baths, resulting in dispersions containing TaSe$_3$ "threads" of approximately 30 to 80 nm wide. Figure 1 (c) shows a high-resolution transmission electron microscopy (HRTEM) image of a TaSe$_3$ thread produced by exfoliation in ethanol. The clearly resolved lattice fringes with 0.808 nm separation are consistent with the (1 0 -1) interplanar distances of TaSe$_3$. As illustrated in the Figure 1 (c) inset, this plane occupies the van der Waals gap of the material, exactly between adjacent TaSe$_3$ sheets. In addition, the HRTEM data shows no obvious structural defects. The electron diffraction patterns also confirm the crystallinity and orientation of the samples (see Supplementary Materials).

Micro-Raman spectroscopy further confirmed the quality of exfoliated TaSe$_3$. Raman measurements were performed in the backscattering configuration under λ = 633 nm laser excitation. Figure 1 (e) shows the Raman spectrum of a TaSe$_3$ thread under small-power excitation (P≤0.5 mW on the surface) in order to avoid local heating. Although the published Raman data for bulk TaSe$_3$ is limited [34] we were able to identify the main Raman peaks. The spectrum displays seven characteristic peaks in the 100 cm$^{-1}$ to 300 cm$^{-1}$ range that originate from the structure of primitive monoclinic TaSe$_3$ [31, 34, 35]. The peaks at 140 cm$^{-1}$, 164 cm$^{-1}$, 214 cm$^{-1}$ and 238 cm$^{-1}$ are attributed to the out-of-plane (A$_{1g}$) modes whereas the peaks at 176 cm$^{-1}$ and 185 cm$^{-1}$ correspond to B$_2$/A$_g$ vibrational modes, with B$_2$ and A$_g$ designating vibrational symmetry in the chain and crystal respectively [34]. The peak at 127 cm$^{-1}$ corresponds to the shear (B$_g$-like) vibration mode [34].

The overall analysis of the Raman, HRTEM and diffraction data indicates that exfoliated TaSe$_3$ is single crystalline, as expected from samples derived from single crystalline CVT-grown material. Each TaSe$_3$ thread consists of bundles of quasi-1D atomic-scale threads based on the hierarchical structure of TaSe$_3$. Importantly, this complex structure disfavors the formation of grain structures within the thread, in contrast to the situation for conventional metals like copper with *fcc* structures [36-40]. This feature is an important factor in downscaling the interconnect wires to few-nm lateral cross-sections.

The TaSe$_3$ threads produced by chemical exfoliation were thinned further by mechanical exfoliation in a fashion similar to that typically used for graphene and MX$_2$ materials [1, 41, 42]. These TaSe$_3$ samples with very high aspect ratios were placed on Si/SiO$_2$ substrates with pre-fabricated metal electrodes for further characterization. Figure 2 (a) shows a representative SEM image of a millimeter long TaSe$_3$ nanowire with the lateral dimensions on the order of 50 nm placed on Ti/Au metal contacts. One can see in this image that the roughness of metal contacts is larger than that of TaSe$_3$ nanowire. The thickness and roughness of the resulting nanowires were accurately determined via atomic force microscopy (AFM). The AFM inspection was used to examine the location and nature of the breakdown points after reaching J$_B$. In Figure 2 (b) one can see an AFM image of the nanowire region with the current induced damage. It is clearly visible that the nanowire consists of multiple parallel crystalline threads.

[Figure 2 (a-b): SEM and AFM of nanowires]

The devices for testing the current-carrying capacity of quasi-1D/quasi-2D TaSe$_3$/$_h$-BN nanowire heterostructures were fabricated from exfoliated TaSe$_3$ with lateral cross-sections in the range 10 nm × 30 nm to 250 nm × 500 nm. Although it was possible to exfoliate nanowires with smaller thicknesses and widths, the selected samples allowed for fabrication of better-quality metal contacts. To protect the exfoliated TaSe$_3$ nanowires from oxidation, we used a $h$-BN capping layer positioned on top of the quasi-1D channels. The capped TaSe$_3$ nanowires were spin coated with the positive resist polymethyl methacrylate (PMMA) and heated two times. To expose the capped TaSe$_3$ nanowires, the assembled structures were selectively etched with sulfur hexafluoride (SF$_6$) gas on an inductively coupled plasma system (Oxford Plasmalab). The metal leads were deposited by electron beam evaporation (Temescal BJD). We used different combinations of metals — 10 nm (Cr, Ti, Au, Pd) / 150 nm Au — in attempts to further improve the contacts. No major difference in performance with different metal contacts was observed. Figure 3 (a) shows the schematic of the fabricated devices. Figure 3 (b) presents an optical image of a representative device – prototype TaSe$_3$ interconnect with h-BN capping layer.

[Figure 3 (a-b): Schematic and optical microscopy of heterostructure devices]

In our heterostructure design, $h$-BN, in addition to providing protection from oxidation, also facilitates heat dissipation from the quasi-1D channels owing to its high thermal conductivity, $K$ [43-45]. There is significant discrepancy in the reported thermal data for h-BN bulk and thin films. The most commonly cited value is $K \approx 600$ W/mK along the basal plane at room temperature (RT) [45]. This is substantially higher than the values for typical TMCs with MX$_3$ atomic composition, which are in the 3.5 – 30 W/mK range [46-49]. The SiO$_2$ layer under TaSe$_3$ is also thermally resistive with the thermal conductivity $K = 0.5-1.4$ W/mK at RT [50]. This is more than one hundred times smaller than that of Si, which has $K = 145$ W/mK in bulk form. For this reason, addition of the *h*-BN capping layer in direct contact with TaSe$_3$ and metal electrodes create an effective heat escape channel. The reduction of the TaSe$_3$ nanowire temperature should increase the breakdown current density irrespective of its mechanism, i.e. electromigration or thermal.

Together with quasi-1D/quasi-2D TaSe$_3$/*h*-BN heterostructures, we fabricated several devices without h-BN capping. Some TaSe$_3$ samples had larger width in order to test possible differences in the quality of metal contacts and $J_B$ values. The device structures for the transmission line measurements (TLM) of the contact resistances were also made. Overall, we tested about 50 devices to ensure reproducibility. The devices with *h*-BN capping had better Ohmic contacts. Figure 4 (a-c) presents the low-field current-voltage characteristics (I-Vs), TLM resistance data and high-field I-Vs near the breakdown point, respectively. One can see from Figure 4 (a) that the contacts are Ohmic. The contact resistance extracted from TLM data is $2R_C=22$ Ω-μm (see Figure 4 (b)). The resistivity is $2.6 - 6.4 \times 10^{-4}$ Ω-cm. This is consisted with the reported data for bulk TaSe$_3$ [31]. Figure 4 (c) shows the high-field I-Vs with the breakdown current density $J_B=32$ MA/cm$^2$. In this specific device the breakdown was gradual with only one critical point.

[Figure 4 (a-c): I-V Characteristics]

Table I presents a representative summary of the breakdown current densities in some of the tested devices. The devices listed in this Table had the channel length $L$ in the range from 2 μm to 13 μm. The surface roughness of as fabricated (not subjected to polishing) TaSe$_3$ channels determined via AFM inspection was in the range from ~0.2 nm to ~0.5 nm. The main conclusion from these data is that $J_B$ values for quasi-1D TaSe$_3$ nanowires capped with quasi-2D h-BN layers can exceed 10 MA/cm$^2$. The measured $J_B$=32 MA/cm$^2$ for a capped device with the thickness $H$=20 nm is a factor of 18 higher than that in state-of-the-art Cu interconnect technology. The *h*-BN capped devices typically had higher $J_B$. The uncapped TaSe$_3$ device with $J_B$=10 MA/cm$^2$ had the smoothest surface with the roughness of ~0.2 nm. The increase in the width of TaSe$_3$ channel (changed $H/W$ ratio) or the use of different metals did not produce major differences in the current-carrying capacity of the nanowires.

**Table I:** Breakdown current density in TaSe$_3$ nanowires and TaSe$_3$/h-BN heterostructures

| Sample | Width (nm) | Thickness (nm) | Current Density (MA/cm$^2$) | Comments |
|---|---|---|---|---|
| A1 | 110 | 20 | 32 | BN capped |
| A2 | 320 | 80 | 11 | BN capped |
| B1 | 75 | 34 | 10 | No capping |
| B2 | 75 | 62 | 4 | No capping |
| B3 | 75 | 40 | 6 | BN Capped |

In approximately half of all devices we observed an unusual type of breakdown: an abrupt step-like decrease of the current (see Figure 5 (a-b)). Since different metal contacts have been used, e.g. Cr/Au and pure Au, it is clear that the step-like breakdown is intrinsic to the quasi-1D TaSe$_3$ nanowires. We attributed this type of breakdown to the *atomic thread* bundle structure of the nanowires. The atomic threads break down one by one or group by group. The result is analogous to several conducting channels in parallel. After the first break, the current abruptly goes down, then starts increasing again with the increasing voltage until the next atomic thread or bundle breaks (see Figure 5 (a)). Interestingly, in a few devices we observed the abrupt increase in the current after the first break, which cannot be explained by increased voltage (see Figure 5 (b)). This is likely to be some sort of "self-repairing" action of the material when individual threads reconnect at high current density regime. The exact mechanism of such repair is reserved for future investigation.

[Figure 5 (a-b): Breakdown I-Vs]

These experiments have demonstrated the promise of quasi-1D TaSe$_3$ in terms of breakdown current density. In order to assess potential of the quasi-1D metals for any interconnect application, we estimate the electrical resistivity of Cu wires with nanometer-scale cross-sections. The electrical resistivity of Cu nanowires increases with the decreasing cross-section due to the electron scatterings on grains and nanowire boundaries. Using the Fuchs-Sondheimer model for the electron–nanowire surface scattering and the Mayadas-Shatzkes model for the electron–grain boundary scattering, the electrical resistivity of metallic nanowires can be written as [33, 51-54]:

$$\rho = \rho_0 \cdot \left[ 2C\lambda_0 \cdot (1-p) \cdot \left( \frac{1}{H} + \frac{1}{W} \right) + \frac{1}{1 - 3\alpha/2 + 3\alpha^2 - 3\alpha^3 \ln(1 + 1/\alpha)} \right] \quad (1)$$

where $\rho_0$ is the bulk electrical resistivity, $W$ is the nanowire width, $H$ is the nanowire thickness, $\lambda_0$ is the bulk electron mean free path (MFP), $C = 1.2$ is the constant for rectangular nanowire [54], $p$ is the specularity parameter determining strength of the electron – nanowire surface scattering, $\alpha = \lambda_0 \cdot R/(d_G \cdot (1-R))$, $R$ is the reflectivity parameter of the electron – grain boundary scattering and $d_G$ is the average grain size. The first term in Eq. (1) describes the electron scattering on nanowire surface roughness while the second term corresponds to the electron scattering on grains. The values of the empirical parameters $p$ and $R$ change between 0 and 1. The specularity parameter $p = 0$ means pure diffusive electron scattering on the nanowire surfaces with maximum resistance to electron transport while $p = 1$ corresponds to the pure specular scattering of electrons without resistance to electron transport. In this model, $R = 1$ means the strongest scattering of electrons on grains without reflection and $R = 0$ corresponds to the total reflection of electrons without scattering.

Figure 6 shows the calculated electrical resistivity of Cu nanowires normalized to the bulk resistivity $\rho_0$ as a function of the nanowire width $W$ for the case $H=W$. The results are shown for a range of parameters *p* and *R* reasonable for Cu [33]. As one can see, the electrical resistivity of Cu nanowires with *W*<10 nm can increase by a factor of 50 to 300 compared to its bulk value. The resistivity increase by a factor of 100 makes it comparable with the values measured for quasi-1D TaSe$_3$ in this work. Because relevant material parameters for TaSe$_3$ are not known, we cannot provide a direct comparison of resistivity scaling between Cu and TaSe$_3$. However, it is reasonable to expect the slower increase in the resistivity of quasi-1D der Waals metals with decreasing cross-section area. In addition, because HRTEM and diffraction data indicate that the TaSe$_3$ samples examined in this work are single crystalline rather than polycrystalline, the possibility of electron scattering on grain boundaries is eliminated (see Eq. (1)). The van der Waals nature of the bonds between the atomic threads should result in much smother boundaries than found in conventional metals.

[Figure 6: Resistivity scaling for Cu]

The electrical current breakdown in thin metallic films is typically of electromigration nature [49]. In some materials, such as carbon nanotubes or graphene, characterized by strong sp$^2$ carbon bonds, the breakdown is thermally induced [50]. There are three types of diffusion processes caused by electromigration: bulk diffusion, grain boundary diffusion and surface diffusion. The grain boundary diffusion is the dominant process in aluminum interconnects whereas surface diffusion limits the breakdown current density in copper interconnects. Since HRTEM and other material characterization studies confirmed the single crystal nature of our TaSe$_3$ nanowires, one can rule out grain boundary diffusion as the failure mechanism. The dominance of the surface diffusion as the breakdown mechanism suggests that reducing the surface roughness of the nanowire, e.g. by chemical or mechanical polishing, can further increase $J_B$. The microscopy inspection of the breakdown points indicate that they are not located at the TaSe$_3$–metal contacts but rather distributed along the length of the nanowire. This means that the electron transport is diffusive rather than ballistic. Improving heat removal with the use of thermally conductive substrates, in addition to *h*-BN capping, can also benefit the current-carrying capacity of quasi-1D van der Waals metals.

In conclusion, we investigated the breakdown current density, $J_B$, of TaSe$_3$/*h*-BN quasi-1D/quasi-2D nanowire heterostructures. It was established that quasi-1D TaSe$_3$ nanowires have $J_B$ exceeding 10 MA/cm$^2$ — an order-of-magnitude higher than that in Cu. The quasi-1D single crystal nature of TaSe$_3$ results in the absence of grain boundaries, low surface roughness, and it potentially can allow for extreme downscaling to the few-nm range, enabling downscaled local interconnects.

## *METHODS*

**Material Preparation:** TaSe$_3$ was synthesized directly from the elements. Tantalum (12.0 mmol, STREM 99.98% purity) and selenium (35.9 mmol, STREM 99.99% purity) were first mixed together. Then iodine (~6.45 mg/cm$^3$, J.T. Baker, 99.9% purity), followed by the tantalum + selenium mixture, was placed in a 17.78 x 1 cm fused quartz ampule (cleaned overnight in a nitric acid soak followed by 12 h anneal at 900 °C). The ampule was evacuated and backfilled with Ar three times while submerged in an acetonitrile/dry ice bath. After flame sealing, ampules were placed in a Carbolite EZS 12/450B three-zone horizontal tube furnace heated at 20 °C min$^{-1}$ to a final temperature gradient of 700 – 680 °C (hot zone – cool zone). The reaction was held at these temperatures for two weeks, then allowed to cool to room temperature. The TaSe$_3$ crystals were removed from the ampule and any remaining I$_2$ removed by sublimation under vacuum. Isolated yield of silver-black crystals was 90.8%. For chemical exfoliation, 6 mg of bulk powdered TaSe$_3$ crystals were sonicated in 10 mL ethanol for several hours. This resulted in a brownish-black non-transparent mixture. The mixture was centrifuged at 2600 rpm for 15 min to remove the larger particles. The remaining dispersed TaSe$_3$ threads were 30 to 80 nm wide.

**Acknowledgements**

The device fabrication work and part of the material characterization conducted at UCR were supported, in part, by the Semiconductor Research Corporation (SRC) and Defense Advanced Research Project Agency (DARPA) through STARnet Center for Function Accelerated nanoMaterial Engineering (FAME). Material synthesis at UGA and part of the material characterization conducted at UGA and UCR were supported by the National Science Foundation's Emerging Frontiers of Research Initiative (EFRI) 2-DARE project: Novel Switching Phenomena in Atomic $MX_2$ Heterostructures for Multifunctional Applications (NSF 005400). The authors thank Edward Hernandez (UCR) for his help with the schematic of the device.

**Contributions**

A.A.B. conceived the idea, coordinated the project, contributed to experimental data analysis and wrote the manuscript; T.T.S. supervised material synthesis and contributed to materials analysis; G.L., M.S., and R.S. performed mechanical exfoliation, fabricated and tested devices; M.A.B. synthesized $TaSe_3$ and conducted materials characterization; E.A. performed chemical exfoliation, Raman spectroscopy and TEM analysis; C.J. performed exfoliation and AFM study; K.N.B. assisted with TEM measurements and data analysis; D.L.N. conducted numerical simulations. All authors contributed to writing of the manuscript.

**Supplementary Information:**

Details of the material synthesis, device fabrication, transmission electron microscopy and diffraction are available on the journal web-site for free-of-charge.

**FIGURES**

**Figure 1**: (a) Crystal structure of monoclinic TaSe$_3$, with alternating corrugated layers of TaSe$_3$ colored blue and yellow. The top view shows the cross section of the unit cell, perpendicular to the chain axis (b axis), which highlights the van der Waals gaps within the material. The side view in the bottom panel emphasizes the 1D nature of TaSe$_3$ chains along the b axis. (b) SEM image of a TaSe$_3$ crystal used in this work. (c) HRTEM image of TaSe$_3$ after solvent exfoliation. The inset shows the position of the observed (1 0 -1) lattice plane in the van der Waals gap. (d) Powder XRD pattern of TaSe$_3$ crystals; the experimental data in black matches the reference pattern in blue (JCPDS 04-007-1143). The intensities of peaks marked with * are enhanced due to orientation effects. (e) Raman spectrum of TaSe$_3$ threads under 633 nm laser excitation.

**Figure 2**: (a) SEM image of a long TaSe$_3$ nanowire with the lateral dimensions on the order of 50 nm placed on top of the Ti/Au metal contacts. Note the much rougher surface of the conventional metal compared to the exfoliated TaSe$_3$ nanowire. (b) Atomic force microscopy image of the breakdown point in the TaSe$_3$ nanowire showing that the nanowires consist of multiple parallel crystalline threads.

**Figure 3**: (a) Schematic of the TaSe$_3$/h-BN quasi-1D/quasi-2D nanowire heterostructures used for the I-V and breakdown current density testing. (b) Optical microscopy images of two fabricated *h*-BN capped TaSe$_3$ nanowire samples. The pseudo colors are used for clarity. The metals tested for fabrication of Ohmic contacts included various combinations of thin layers of Cr, Ti, Au, Pd together with a thicker Au layer.

**Figure 4**: (a) Low-field I-V characteristics of TaSe$_3$/*h*-BN quasi-1D/quasi-2D nanowire heterostructures indicating that the contacts are Ohmic. (b) Transmission line measurement resistance data. (c) High-field I-V characteristics showing the breakdown point. In this specific device the breakdown is gradual. The inset shows a SEM image of the TLM structure used in the low-field testing.

**Figure 5**: (a) Current-voltage characteristics of a device with Cr/Au (10/150 nm) contacts. Note a step-like breakdown starting at $J_B = 4\times10^6$ A/cm$^2$. (b) Current-voltage characteristics of devices with pure Au contacts (150 nm) showing the step-like breakdowns at $J_B = 6.1\times10^6$ A/cm$^2$ (black), $5.7\times10^6$ A/cm$^2$ (blue), and $6.3\times10^6$ A/cm$^2$ (red). The observed step-like breakdown was attributed to the quasi-1D atomic thread crystal structure of the material. The inset shows a microscopy image of a representative *h*-BN capped TaSe$_3$ nanowire device used in testing. The blue colored region in the image is *h*-BN layer.

**Figure 6**: Calculated electrical resistivity of Cu nanowires normalized to the bulk Cu resistivity as a function of the nanowire width *W*. The results are shown for a range of specularity parameters *p*, which defines electron scattering from nanowire surfaces and parameter *R*, which determines the electron scattering from grain boundaries. Note a strong increase in Cu resistivity as the lateral dimensions approach a few-nm range. The increase in TaSe$_3$ resistivity is expected to be less drastic owing to the absence of grain boundaries ($R\rightarrow0$) and smother surfaces ($p\rightarrow1$).

# REFERENCES

1. Novoselov, K.S.; Geim, A.K.; Morozov, S.V.; Jiang, D.; Zhang, Y.; Dubonos, S.V.; Grigorieva, I.V.; Firsov, A.A., *Science*, **2004**, *306*(5696), 666-669.
2. Zhang, Y.B.; Tan, Y.W.; Stormer, H.L.; Kim, P., *Nature*, **2005**, *438*(7065), 201-204.
3. Balandin, A.A.; Ghosh, S.; Bao, W.Z.; Calizo, I.; Teweldebrhan, D.; Miao, F.; Lau, C.N., *Nano Letters*, **2008**, *8*(3), 902-907.
4. Balandin, A.A., *Nature Materials*, **2011**, *10*(8), 569-581.
5. Casiraghi, C.; Hartschuh, A.; Lidorikis, E.; Qian, H.; Harutyunyan, H.; Gokus, T.; Novoselov, K.S.; Ferrari, A.C., *Nano Letters*, **2007**, *7*(9), 2711-2717.
6. Nair, R.R.; Blake, P.; Grigorenko, A.N.; Novoselov, K.S.; Booth, T.J.; Stauber, T.; Peres, N.M.R.; Geim, A.K., *Science*, **2008**, *320*(5881), 1308-1308.
7. Mak, K.F.; Shan, J.; Heinz, T.F., *Physical Review Letters*, **2011**, *106*(4).
8. Novoselov, K.S.; Jiang, D.; Schedin, F.; Booth, T.J.; Khotkevich, V.V.; Morozov, S.V.; Geim, A.K., *Proceedings of the National Academy of Sciences of the United States of America*, **2005**, *102*(30), 10451-10453.
9. Bolotin, K.I.; Sikes, K.J.; Jiang, Z.; Klima, M.; Fudenberg, G.; Hone, J.; Kim, P.; Stormer, H.L., *Solid State Communications*, **2008**, *146*(9-10), 351-355.
10. Cui, X.; Lee, G.H.; Kim, Y.D.; Arefe, G.; Huang, P.Y.; Lee, C.H.; Chenet, D.A.; Zhang, X.; Wang, L.; Ye, F.; Pizzocchero, F.; Jessen, B.S.; Watanabe, K.; Taniguchi, T.; Muller, D.A.; Low, T.; Kim, P.; Hone, J., *Nature Nanotechnology*, **2015**, *10*(6), 534-540.
11. Radisavljevic, B.; Radenovic, A.; Brivio, J.; Giacometti, V.; Kis, A., *Nature Nanotechnology*, **2011**, *6*(3), 147-150.
12. Geim, A.K.; Grigorieva, I.V., *Nature*, **2013**, *499*(7459), 419-425.
13. Teweldebrhan, D.; Goyal, V.; Balandin, A.A., *Nano Letters*, **2010**, *10*(4), 1209-1218.
14. Goli, P.; Khan, J.; Wickramaratne, D.; Lake, R.K.; Balandin, A.A., *Nano Letters*, **2012**, *12*(11), 5941-5945.
15. Balandin, A.A., *Nature Nanotechnology*, **2013**, *8*(8), 549-555.
16. Stolyarov, M.A.; Liu, G.X.; Rumyantsev, S.L.; Shur, M.; Balandin, A.A., *Applied Physics Letters*, **2015**, *107*(2).
17. Mattheis.Lf, *Physical Review B*, **1973**, *8*(8), 3719-3740.
18. Wilson, J.A.; Yoffe, A.D., *Advances in Physics*, **1969**, *18*(73), 193-&.
19. Kuc, A.; Zibouche, N.; Heine, T., *Physical Review B*, **2011**, *83*(24).
20. Mak, K.F.; Lee, C.; Hone, J.; Shan, J.; Heinz, T.F., *Physical Review Letters*, **2010**, *105*(13).
21. Liu, L.T.; Kumar, S.B.; Ouyang, Y.; Guo, J., *Ieee Transactions on Electron Devices*, **2011**, *58*(9), 3042-3047.
22. Ding, Y.; Wang, Y.L.; Ni, J.; Shi, L.; Shi, S.Q.; Tang, W.H., *Physica B-Condensed Matter*, **2011**, *406*(11), 2254-2260.
23. Kam, K.K.; Parkinson, B.A., *Journal of Physical Chemistry*, **1982**, *86*(4), 463-467.
24. Beal, A.R.; Hughes, H.P.; Liang, W.Y., *Journal of Physics C-Solid State Physics*, **1975**, *8*(24), 4236-4248.
25. Wilson, J.A.; Disalvo, F.J.; Mahajan, S., *Advances in Physics*, **1975**, *24*(2), 117-201.
26. Yan, Z.; Jiang, C.; Pope, T.R.; Tsang, C.F.; Stickney, J.L.; Goli, P.; Renteria, J.; Salguero, T.T.; Balandin, A.A., *Journal of Applied Physics*, **2013**, *114*(20).
27. Samnakay, R.; Wickramaratne, D.; Pope, T.R.; Lake, R.K.; Salguero, T.T.; Balandin, A.A., *Nano Letters*, **2015**, *15*(5), 2965-2973.
28. Lipatov, A.; Wilson, P.M.; Shekhirev, M.; Teeter, J.D.; Netusil, R.; Sinitskii, A., *Nanoscale*, **2015**, *7*(29),

12291-12296.
29. Island, J.O.; Buscema, M.; Barawi, M.; Clamagirand, J.M.; Ares, J.R.; Sanchez, C.; Ferrer, I.J.; Steele, G.A.; van der Zant, H.S.J.; Castellanos-Gomez, A., *Advanced Optical Materials*, **2014**, *2*(7), 641-645.
30. Island, J.O.; Barawi, M.; Biele, R.; Almazan, A.; Clamagirand, J.M.; Ares, J.R.; Sanchez, C.; van der Zant, H.S.J.; Alvarez, J.V.; D'Agosta, R.; Ferrer, I.J.; Castellanos-Gomez, A., *Advanced Materials*, **2015**, *27*(16), 2595-2601.
31. Bjerkelund, E.; Fermor, J.H.; Kjekshus, A., *Acta Chemica Scandinavica*, **1966**, *20*(7), 1836-&.
32. *International Technology Roadmap for Semiconductors*, **2013**.
33. Steinhogl, W.; Schindler, G.; Steinlesberger, G.; Engelhardt, M., *Physical Review B*, **2002**, *66*(7).
34. Wieting, T.J.; Grisel, A.; Levy, F., *Molecular Crystals and Liquid Crystals*, **1982**, *81*(1-4), 835-842.
35. Bjerkelund, E.; Kjekshus, A., *Acta Chemica Scandinavica*, **1965**, *19*(3), 701-&.
36. Gertsman, V.Y.; Hoffmann, M.; Gleiter, H.; Birringer, R., *Acta Metallurgica Et Materialia*, **1994**, *42*(10), 3539-3544.
37. Yang, J.J.; Huang, Y.L.; Xu, K.W., *Surface & Coatings Technology*, **2007**, *201*(9-11), 5574-5577.
38. Qiu, T.Q.; Tien, C.L., *Journal of Heat Transfer-Transactions of the Asme*, **1993**, *115*(4), 842-847.
39. Zielinski, E.M.; Vinci, R.P.; Bravman, J.C., *Journal of Applied Physics*, **1994**, *76*(8), 4516-4522.
40. Goli, P.; Ning, H.; Li, X.S.; Lu, C.Y.; Novoselov, K.S.; Balandin, A.A., *Nano Letters*, **2014**, *14*(3), 1497-1503.
41. Dean, C.R.; Young, A.F.; Meric, I.; Lee, C.; Wang, L.; Sorgenfrei, S.; Watanabe, K.; Taniguchi, T.; Kim, P.; Shepard, K.L.; Hone, J., *Nature Nanotechnology*, **2010**, *5*(10), 722-726.
42. Koski, K.J.; Cui, Y., *Acs Nano*, **2013**, *7*(5), 3739-3743.
43. Sichel, E.K.; Miller, R.E.; Abrahams, M.S.; Buiocchi, C.J., *Physical Review B*, **1976**, *13*(10), 4607-4611.
44. Jo, I.; Pettes, M.T.; Kim, J.; Watanabe, K.; Taniguchi, T.; Yao, Z.; Shi, L., *Nano Letters*, **2013**, *13*(2), 550-554.
45. Lindsay, L.; Broido, D.A., *Physical Review B*, **2011**, *84*(15).
46. Guilmeau, E.; Berthebaud, D.; Misse, P.R.N.; Hebert, S.; Lebedev, O.I.; Chateigner, D.; Martin, C.; Maignan, A., *Chemistry of Materials*, **2014**, *26*(19), 5585-5591.
47. Inyushkin, A.V.; Taldenkov, A.N.; Florentiev, V.V., *Synthetic Metals*, **1987**, *19*(1-3), 843-848.
48. Brill, J.W.; Tzou, C.P.; Verma, G.; Ong, N.P., *Solid State Communications*, **1981**, *39*(2), 233-237.
49. Zawilski, B.M.; Littleton, R.T.; Lowhorn, N.D.; Tritt, T.M., *Solid State Communications*, **2010**, *150*(29-30), 1299-1302.
50. Yamane, T.; Nagai, N.; Katayama, S.; Todoki, M., *Journal of Applied Physics*, **2002**, *91*(12), 9772-9776.
51. Fuchs, K., *Proceedings of the Cambridge Philosophical Society*, **1938**, *34*, 100-108.
52. Sondheimer, E.H., *Phil. Mag. Suppl.*, **1952**, *1*, 1.
53. Mayadas, A.F.; Shatzkes, M., *Physical Review B*, **1970**, *1*(4), 1382.
54. Huang, Q.; Lilley, C.M.; Bode, M.; Divan, R.S., *8th IEEE conference on nanotecnology*, **2008**, 549-552.

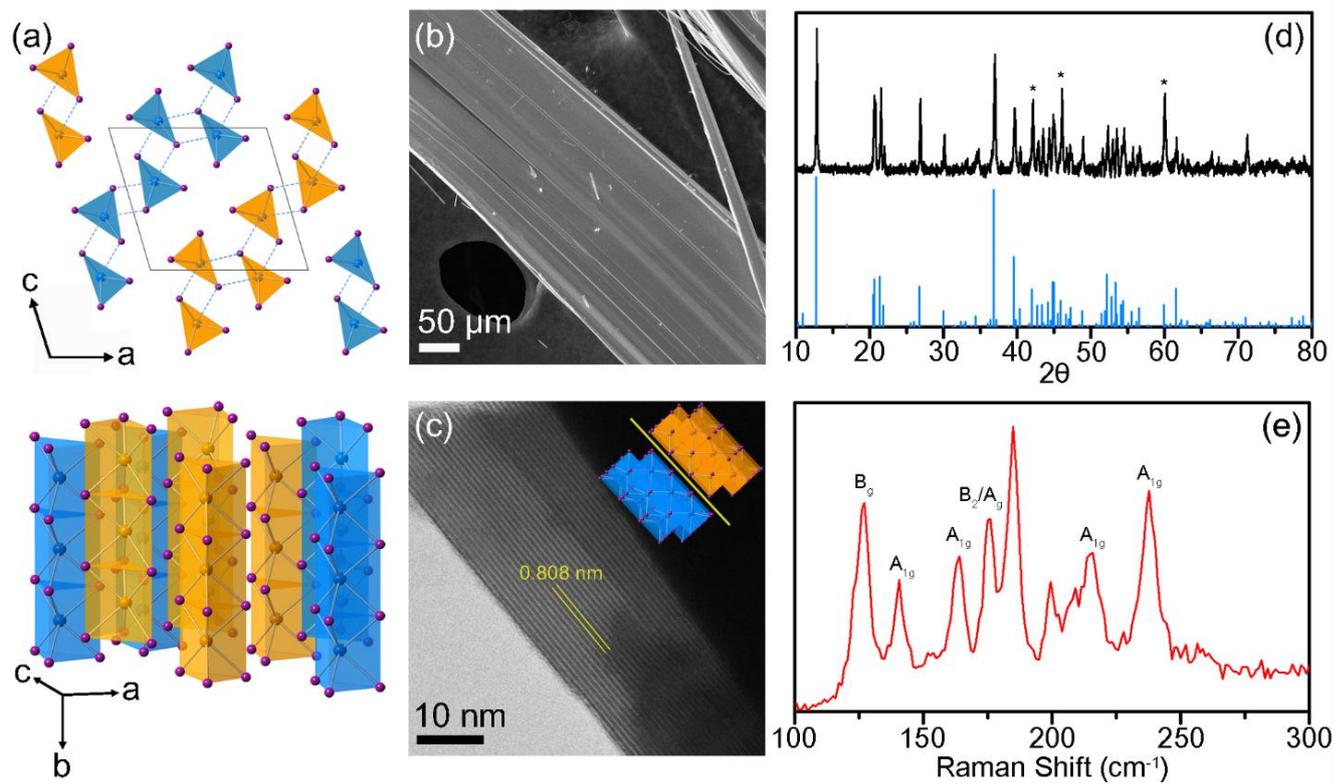

Figure 1

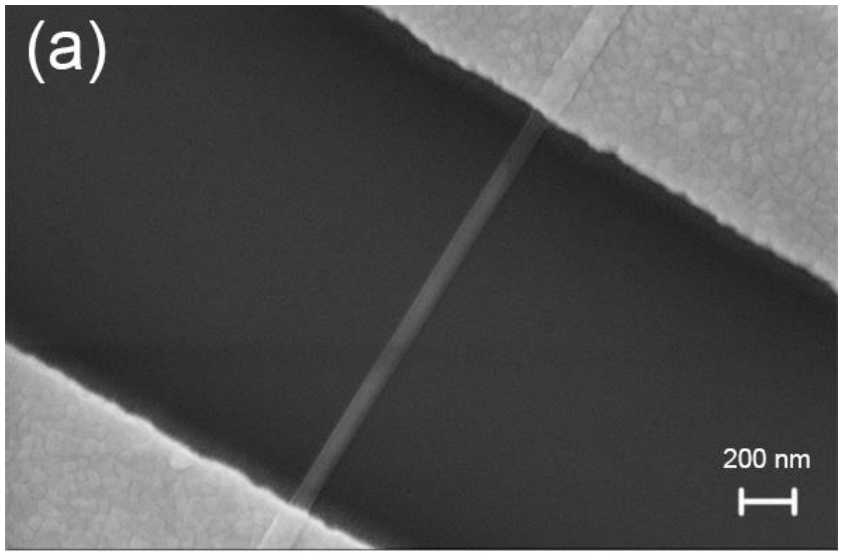

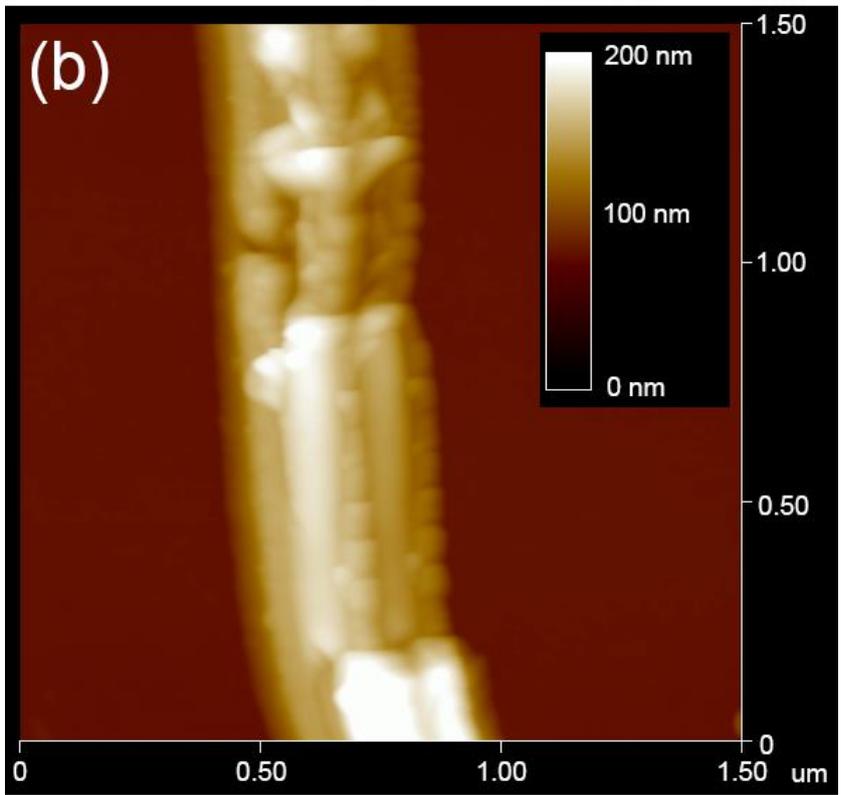

Figure 2

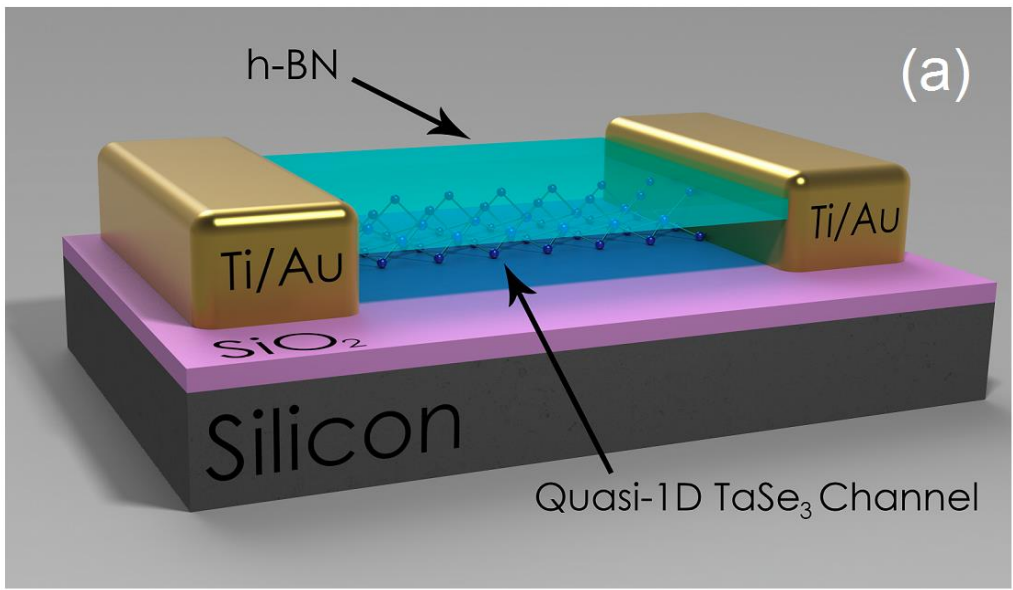

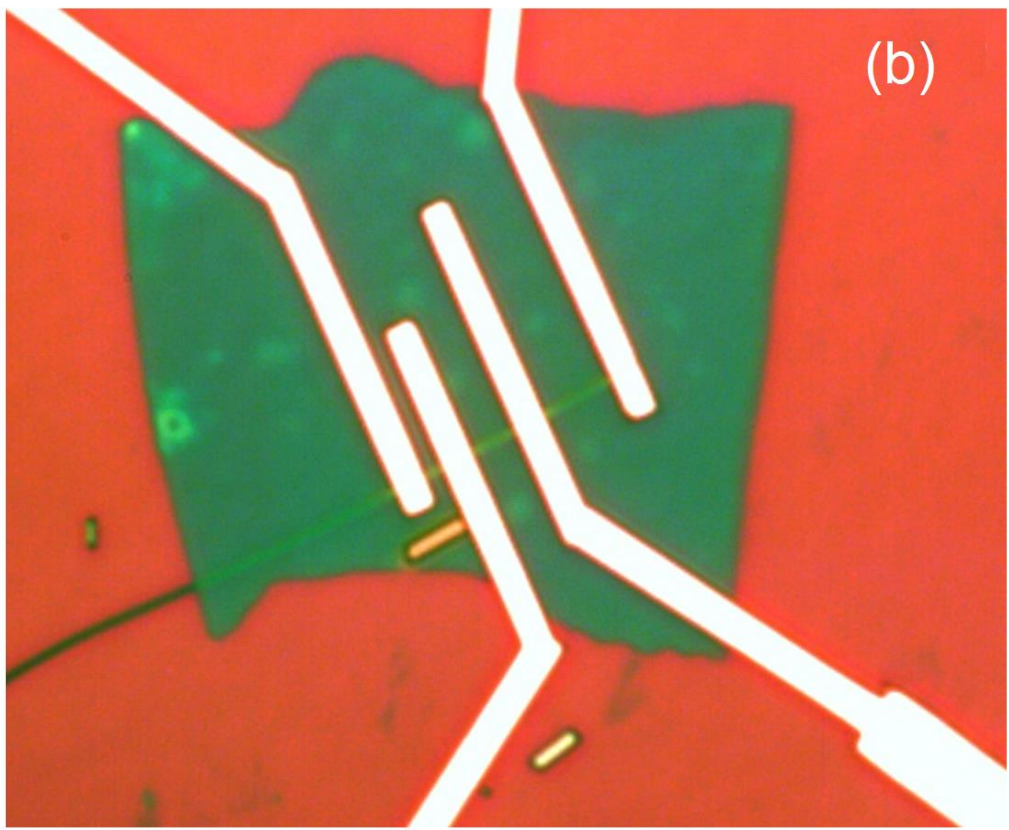

Figure 3

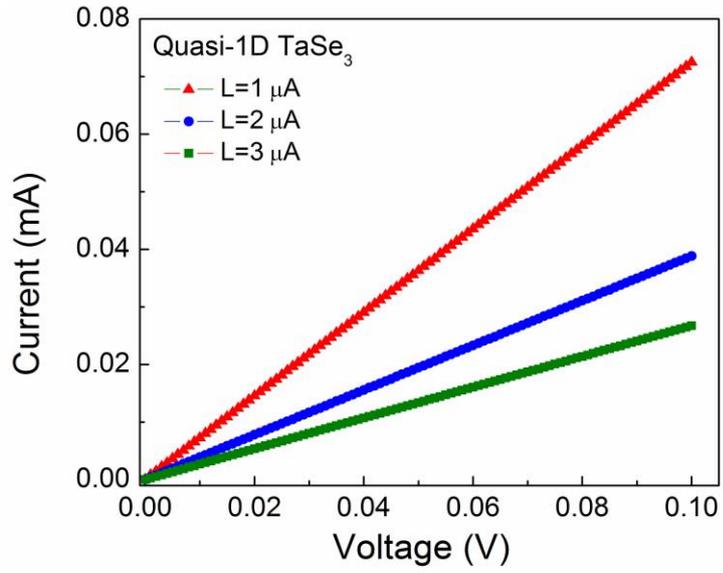

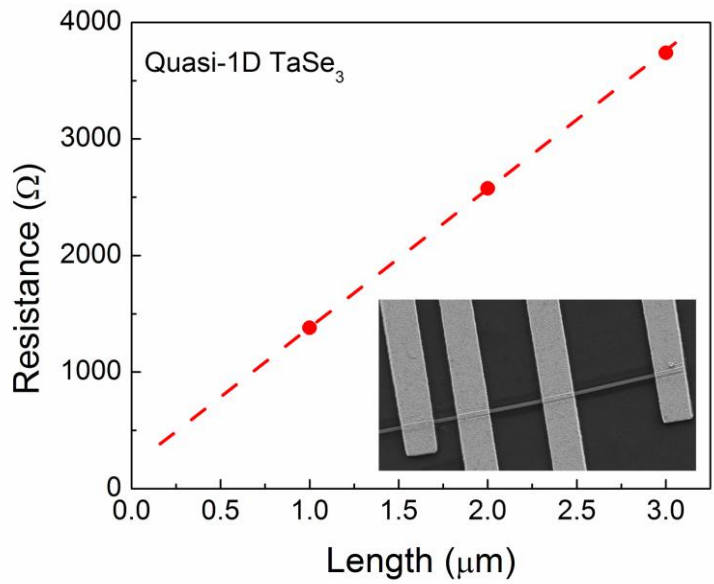

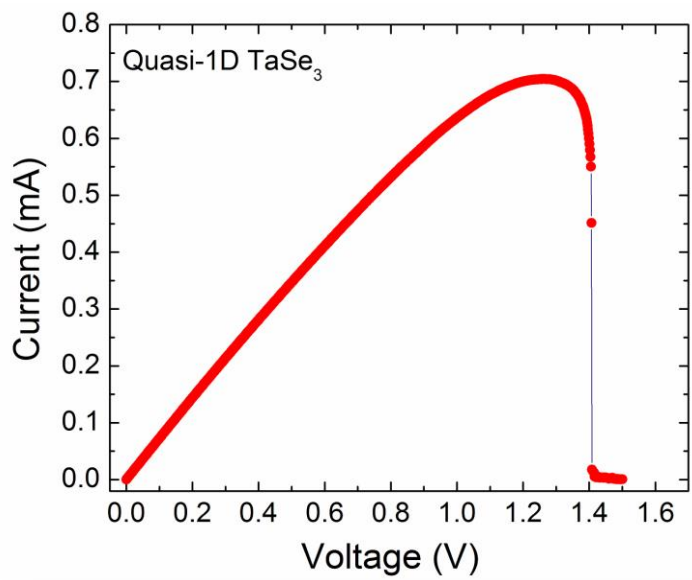

Figure 4

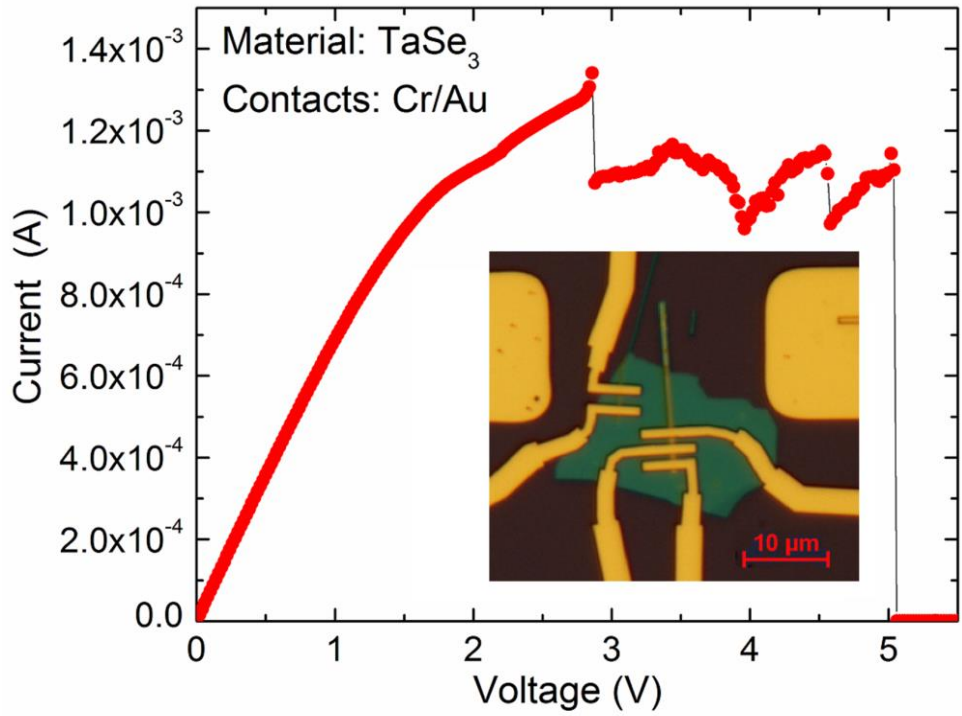
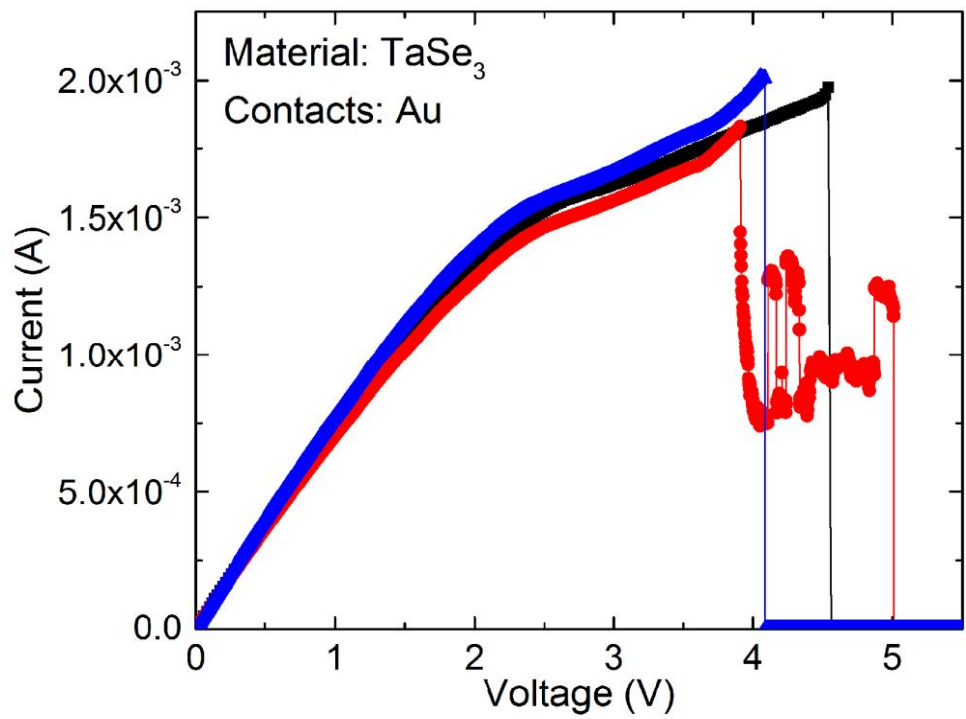

Figure 5

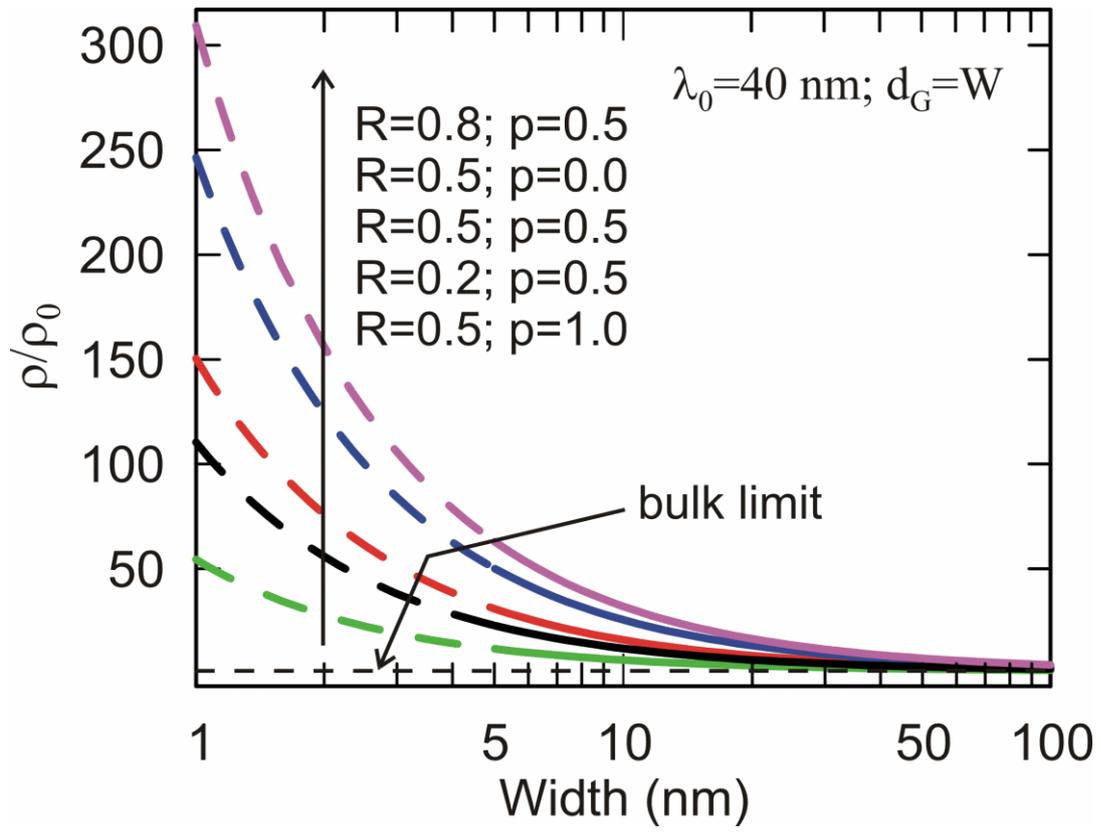

Figure 6